\documentstyle[12pt,epsfig]{article}

\newlength{\dinwidth}
\newlength{\dinmargin}
\def\lsim{\mathrel{\rlap{\lower4pt\hbox{\hskip1pt$\sim$}}
    \raise1pt\hbox{$<$}}}                
\def\gsim{\mathrel{\rlap{\lower4pt\hbox{\hskip1pt$\sim$}}
    \raise1pt\hbox{$>$}}}                

\begin{document}


\begin{flushright}
 ANL-HEP-CP-99-43 \\
 NIKHEF 99-012
\end{flushright}


\begin{center}
\begin{Large}
\begin{bf}

Detailed Comparison of Next-to-Leading Order \\
Predictions for Jet Photoproduction at HERA

\end{bf}
\end{Large}
\vspace*{5mm}
\begin{large}
B.W.\ Harris$^a$, M.\ Klasen$^a$, and J.\ Vossebeld$^{b}$ \\
\end{large}
\end{center}
$^a$ Argonne National Laboratory, Argonne, IL, USA \\
$^b$ NIKHEF, Amsterdam, the Netherlands \\
\begin{quotation}
\noindent
{\bf Abstract:}
The precision of new HERA data on jet photoproduction opens up the possibility
to discriminate between different models of the photon structure. This requires
equally precise theoretical predictions from perturbative QCD calculations.
In the past years, next-to-leading order calculations for
the photoproduction of jets at HERA have become available. Using the kinematic
cuts of recent ZEUS analyses, we compare the predictions of three calculations
for different dijet and three-jet distributions. We find that in general all
three calculations agree within the statistical accuracy of the Monte Carlo
integration yielding reliable theoretical predictions. In certain restricted
regions of phase space, the calculations differ by up to 5\%.
\end{quotation}



\section{Introduction}
\label{sec:1}

Our present knowledge of the hadronic structure of the photon rests on rather
limited data from inclusive deep-inelastic electron-photon scattering. At
leading order (LO) of perturbative QCD, the photon structure function
$F_2^\gamma (x,Q^2)$ is related to the singlet quark densities (dominated by
the up-quark density) which are the only well constrained parton densities in
the photon. In contrast, the gluon density in the photon is only constrained
theoretically by a global momentum sum rule. Experimental constraints are weak,
since the gluon contributes to $F_2^\gamma (x,Q^2)$ only at next-to-leading
order (NLO) of QCD. Therefore, the available parametrizations of the photon
structure function rely heavily on assumptions like Vector Meson Dominance.
Valuable information on the gluon density in the photon is provided by
jet photoproduction, where existing data have already ruled out a very large
and hard gluon density.

Jet photoproduction has been measured with increasing precision at HERA since
the electron-proton collider became operational in 1992. These data make it
now possible to discriminate between different parametrizations of the photon
structure if uncertainties from the proton structure and from the partonic
scattering process can be minimized. The proton structure is well constrained
in the relevant regions of $x$ from deep-inelastic HERA data. Direct and
resolved photon-proton scattering processes into one or two jets have been
calculated by three groups in NLO QCD \cite{Frixione:1997ks,Harris:1997hz,%
Klasen:1996it}. These calculations are also applicable to LO three-jet
production. The purpose of this paper is to check the consistency of these
three calculations using the kinematic cuts of recent ZEUS dijet \cite{dijet}
and three-jet \cite{Breitweg:1998uv} analyses at a precision that is only
limited by the accuracy of the numerical integration. The organization
of the paper
is as follows: In Sect.\ \ref{sec:2} we briefly describe the theoretical
methods used in the perturbative calculations. In Sect.\ \ref{sec:3} we present
a detailed comparison of the LO three-jet distributions, and in Sect.\
\ref{sec:4} we present the comparison of the NLO dijet distributions. In Sect.\
\ref{sec:5} we discuss remaining theoretical and experimental uncertainties,
and we give our conclusions in Sect.\ \ref{sec:6}.



\section{Theoretical Methods Used in the NLO Calculations}
\label{sec:2}

The basic components in current NLO jet photoproduction calculations are
$2\rightarrow 2$ body squared matrix elements through one-loop order and
tree-level $2\rightarrow 3$ body squared matrix elements, for both
photon-parton and parton-parton initiated subprocesses. It is therefore
possible to study single- and dijet production at NLO and three-jet production
at LO. The goal of the next-to-leading order calculations is to organize the
soft and collinear singularity cancellations without loss of information 
in terms of observable quantities.
The methods to accomplish this cancellation can be categorized 
as the phase space slicing and subtraction methods.

The calculation of \cite{Frixione:1997ks} uses the subtraction method.
In the center of mass frame of the incoming parton the final state 
parton four vectors may be written as 
$p_i = \frac{\sqrt{S}}{2} \xi_i (1,\sqrt{1-y_i^2}\vec{e}_{iT},y_i)$
where $\vec{e}_{iT}$ is a transverse unit vector.  By construction, 
the parton $i$ gets soft when $\xi_i \rightarrow 0$, and collinear to 
the incoming partons when $y_i \rightarrow \pm 1$.  The $n$-dimensional
three-body phase space written in terms of $\xi_i$ and $y_i$ is proportional 
to $\xi_i^{1-2\epsilon}(1-y_i^2)^{-\epsilon}$, where $\epsilon=2-n/2$.
The soft singularities in the matrix element squared, which are of
${\cal O}(\xi_i^{-2})$, are regulated 
by multiplying them by $\xi_i^2$ and at the same time dividing the 
phase space by $\xi_i^2$, resulting in a 
$\xi_i^{-1-2\epsilon}(1-y_i^2)^{-\epsilon}$ structure.  
The term $\xi_i^{-1-2\epsilon}$ is replaced by plus distributions 
and soft poles in $\epsilon$.  Within these terms 
$(1-y_i^2)^{-\epsilon}$ is replaced by additional plus 
distributions and collinear poles in $\epsilon$.
The soft and final state collinear singularities cancel upon addition 
of the interference of the leading order diagrams with the 
renormalized one-loop virtual diagrams.
The initial state collinear singularities are removed through mass 
factorization.  
The result is a finite function of 
various combinations of plus distributions.  The main drawback of this 
method is that in the subtracted integrals, the numerical singularity 
cancellation takes place between terms of different kinematics 
which requires special care.

The calculation of \cite{Harris:1997hz} uses a phase space slicing 
method that employs two small cut-offs $\delta_s$ and $\delta_c$ to
delineate soft and collinear regions of phase space.  This avoids 
partial fractioning at the expense of a somewhat more complicated 
split of phase space.
Defining the four vectors of the three-body scattering process as 
$p_1+p_2 \rightarrow p_3+p_4+p_5$ one takes $s_{ij}=(p_i+p_j)^2$.  The 
soft region is then $E_i < \delta_s\sqrt{s_{12}}/2$ where 
$E_i$ is the energy of the emitted gluon.  In this region 
one puts $p_i=0$ everywhere except in denominators of matrix elements, 
and performs the integral over the restricted phase space in $n$ dimensions.
The complementary region, $E_i > \delta_s\sqrt{s_{12}}/2$, is called the 
hard region.  That portion of the hard region satisfying 
$s_{ij} \; {\rm or} \; |t_{ij}| < \delta_c s_{12} $ with $t_{ij}=(p_i-p_j)^2$
is treated with collinear kinematics and is also integrated
in $n$ dimensions.  The poles in $\epsilon$ cancel as described above, and 
terms of order $\delta_c$ and $\delta_s$ are neglected 
compared to double and single logarithms of the cut-offs.  
The hard and non-collinear phase space region is integrated numerically.
The sum is independent of the cut-offs 
provided they are  chosen small enough. This 
serves as a useful check on results.

The calculation of \cite{Klasen:1996it}, which is also of the phase space 
slicing type, uses an invariant mass cut to isolate singular regions of 
phase space.  The $2\rightarrow 3$ body squared matrix elements are partially
fractioned to  separate overlapping soft and collinear singularities.
As above one defines $s_{ij}=(p_i+p_j)^2$.  
In the situation when $s_{ij} \leq ys_{12}$ the partons $i$ and $j$ cannot 
be resolved.  In this region the phase space integrals are performed in $n$ 
dimensions which produces double and single poles in $\epsilon$. 
They cancel as described above.  Terms of order $y$ are neglected in the 
process, but double and single logarithms in $y$ are retained.  
The region $s_{ij} > ys_{12}$ is integrated numerically.  The sum is 
independent of $y$ provided it is chosen small enough. As above, this 
serves as a useful check on results.

The final result of these calculations is an expression 
that is finite in four-dimensional space-time.  One can 
compute all phase space integrations using 
standard Monte-Carlo integration techniques.
The result is a program which returns 
parton kinematic configurations and their corresponding 
weights, accurate to ${\cal O}(\alpha\alpha_s^2)$.  
The user is free to histogram any set of 
infrared-safe observables and apply parton
level cuts, all in a single histogramming subroutine.
The calculations have the added benefit that when one 
considers a manifestly three-body observable the two 
body contributions don't contribute and a leading order three-jet 
prediction results.



\section{Three-Jet Cross Sections}
\label{sec:3}

During 1995 and 1996, positrons of energy $E_e = 27.5$ GeV were collided at
HERA with protons of energy $E_p = 820$ GeV. In ZEUS photoproduction
events were selected
by anti-tagging the positron such that the photon has a virtuality $Q^2$
smaller than 1 GeV$^2$ and an energy fraction in the positron $0.2 < y < 0.8$.
Three-jet events were analysed with a $k_T$ clustering
algorithm using a jet separation
parameter of $R=1$ in a rapidity range of $|\eta| < 2.4$. The jets were
required to have transverse energies above 6 GeV (two highest $E_T$ jets) and
above 5 GeV (third jet). Additional cuts were placed on the three-jet mass
$M_{\rm 3-jet} > 50$ GeV, the leading jet energy fraction $x_3 < 0.95$, and the
cosine of the leading jet scattering angle $|\cos\theta_3| < 0.8$
\cite{Breitweg:1998uv}. NLO calculations for three-jet photoproduction are not
yet available, so the theoretical predictions for three-jet distributions,
which are compared here,
are only accurate to LO. Therefore one tests only the $2\rightarrow 3$ phase
space generators and the tree-level $2\rightarrow 3$ matrix elements, but no
soft or collinear singular regions. All calculations use CTEQ4L
\cite{Lai:1997mg} and GRV-LO \cite{Gluck:1992ee}
parton distributions in the proton and photon, respectively. The strong
coupling constant $\alpha_s(\mu)$ is calculated in leading order with five
flavors and $\Lambda_{\rm QCD}^{(5)} = 181$ MeV, and the renormalization and
factorization scale $\mu$ is identified with the largest transverse energy
of the three jets.

\begin{figure}[ttt]
 \begin{center}
  \vspace*{-1cm}
  \epsfig{file=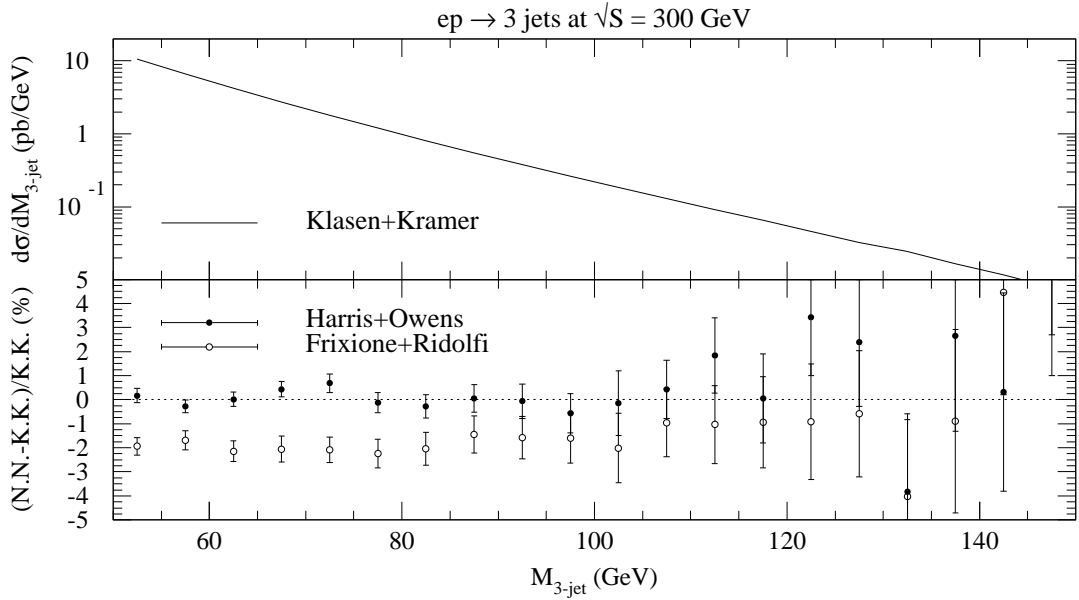,width=16cm}
  \vspace*{-1cm}
 \end{center}
 \caption{Comparison of three theoretical predictions for the LO three-jet
 cross section as a function of the three-jet mass $M_{\rm 3-jet}$. The cross
 section falls exponentially with $M_{\rm 3-jet}$. HO agree
 very well with KK, whereas FR are systematically 2\% lower.}
 \label{fig:m3j}
\end{figure}

In Fig.\ \ref{fig:m3j} we compare the theoretical predictions for the
LO three-jet
mass distribution by Harris and Owens (HO) and by Frixione and Ridolfi (FR)
to those
by Klasen and Kramer (KK). In the upper figure we plot the absolute cross
section which falls exponentially with $M_{\rm 3-jet}$. This demonstrates
that the total cross section is dominated by the region close to $M_{\rm
3-jet} > 50$ GeV. In the lower figure we plot the relative difference between
the results by HO and by FR to the results by
KK, normalized to the latter. The statistical accuracy of the
different calculations is comparable. It has been included in the error bars
and decreases simultaneously with the magnitude of the cross section. The
calculation of HO presented here differs from the previous
results as published in \cite{Breitweg:1998uv}, where the $E_T$ cuts were
applied to energy, not transverse energy, ordered jets. It now agrees very
well (better than 0.5\% at low $M_{\rm 3-jet}$) with that by KK.
The calculation by FR is systematically 2\% lower.

\begin{figure}[ttt]
 \begin{center}
  \vspace*{-1cm}
  \epsfig{file=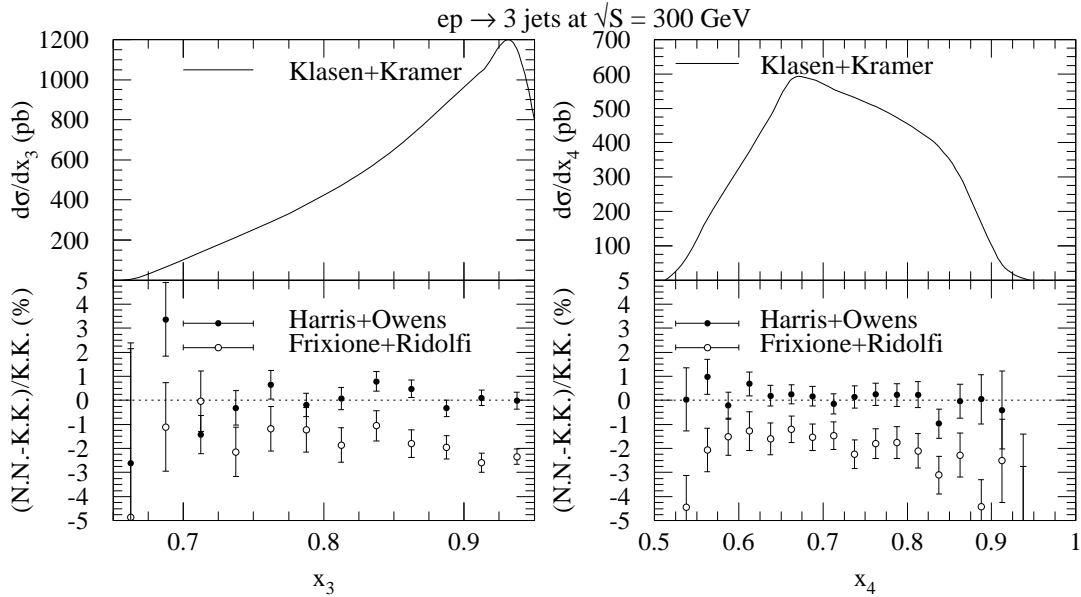,width=16cm}
  \vspace*{-1cm}
 \end{center}
 \caption{Comparison of three theoretical predictions for the LO three-jet
 cross section as a function of the energy fractions $x_3$ (left) and $x_4$
 (right) of the leading and next-to-leading jets. The distributions are
 dominated by the available phase space. HO agree
 very well with KK, whereas FR are
 systematically 2\% lower.}
 \label{fig:x3x4}
\end{figure}

A similar comparison for the distributions in the energy fractions of the
leading and next-to-leading jets is shown in Fig.\ \ref{fig:x3x4}.
These distributions are dominated by the available phase space, not the QCD
dynamics, and thus present a test on the two-to-three phase space
generators of the numerical programs. The statistical accuracy depends again
on the size of the cross section. Where the cross section is large, HO
agree with KK to better than 0.5\%. FR
are again 2\% lower, which indicates that the difference may come from
the phase space generator or kinematic cuts.

\begin{figure}[ttt]
 \begin{center}
  \vspace*{-1cm}
  \epsfig{file=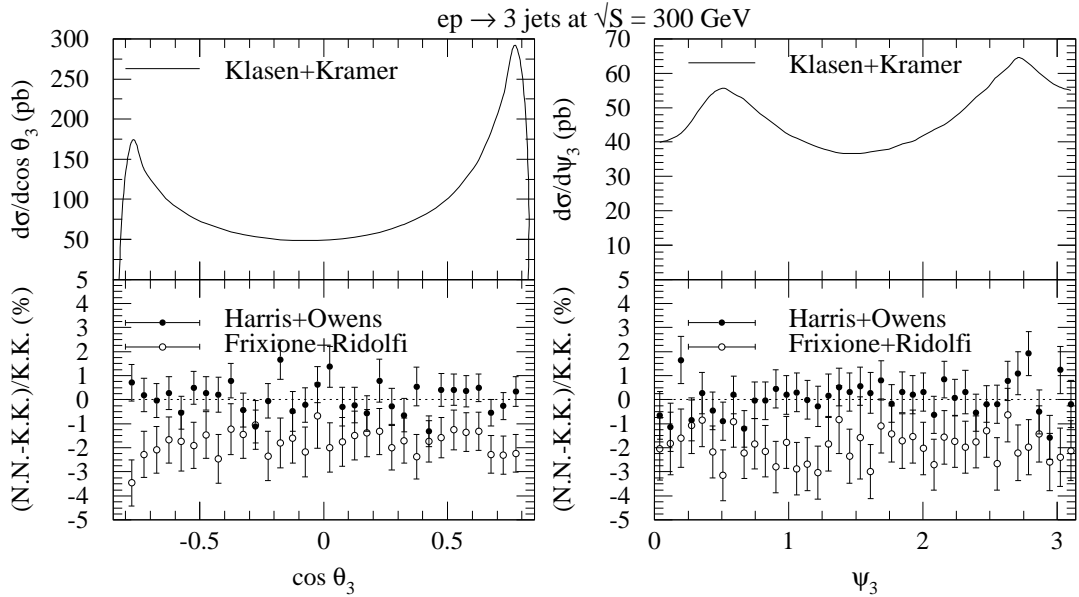,width=16cm}
  \vspace*{-1cm}
 \end{center}
 \caption{Comparison of three theoretical predictions for the LO three-jet
 cross section as a function of the cosine of the fastest jet scattering angle
 $\cos\theta_3$ (left) and the angle $\psi_3$ between the three jet plane and
 the plane containing the leading jet and the average beam direction (right).
 The distributions are sensitive to the pole structure of the QCD matrix
 elements. HO agree
 very well with KK, whereas FR are
 systematically 2\% lower.}
 \label{fig:cost3psi3}
\end{figure}

The QCD matrix elements are tested in distributions of the cosine of the
fastest jet scattering angle $\cos\theta_3$ and the angle $\psi_3$ between the
three-jet plane and the plane containing the leading jet and the average beam
direction. The results are presented in Fig.\ \ref{fig:cost3psi3}.
We find again very good agreement between HO and KK
and a 2\% difference with FR.



\section{Dijet Cross Sections}
\label{sec:4}

For the dijet photoproduction analysis ZEUS selected again photons with a
virtuality below 1 GeV$^2$. The range of the energy fraction of the photon
in the positron $0.2 < y < 0.85$ was slightly larger than in the three-jet
analysis, and in addition a narrower band of $0.5 < y < 0.85$ was analyzed
which enhances the sensitivity to the parton densities in the photon.
The transverse energy of the leading (second) jet was
required to be larger than 14 (11) GeV with both jets lying in the rapidity
region of $-1 < \eta_{1,2} < 2$ \cite{dijet}. All NLO calculations use the
CTEQ4M
and GRV-HO parton densities for the proton and photon and
$\Lambda_{\rm QCD}^{(5)} = 202$ MeV corresponding to CTEQ4M in the NLO
approximation of $\alpha_s(\mu=\max(E_{T_{1,2}}))$.

\begin{figure}[ttt]
 \begin{center}
  \vspace*{-1cm}
  \epsfig{file=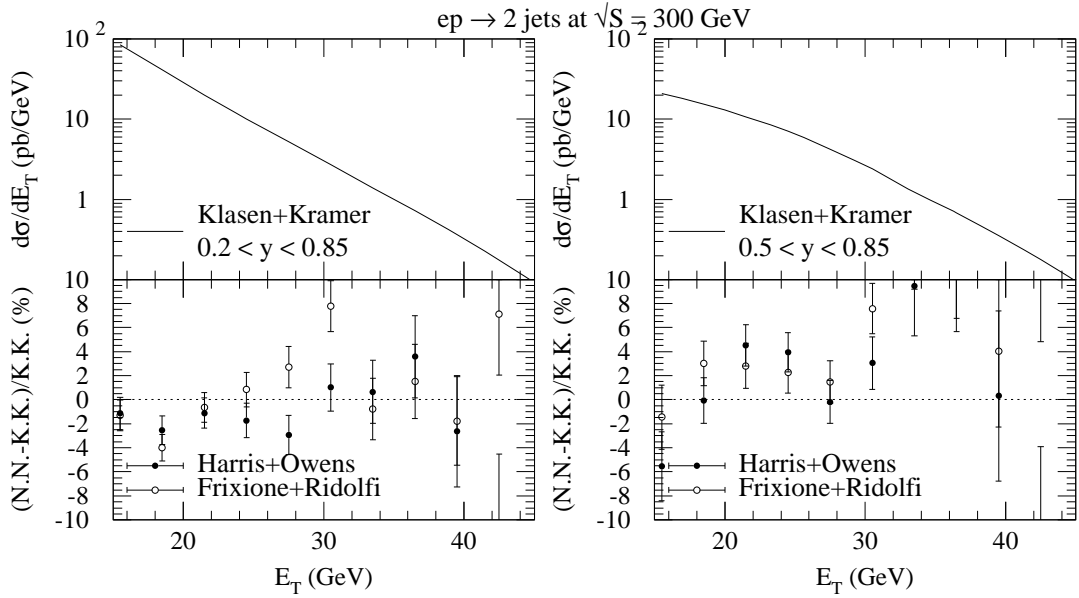,width=16cm}
  \vspace*{-1cm}
 \end{center}
 \caption{Comparison of three theoretical predictions for the NLO dijet
 cross section as a function of the transverse energy $E_T$ of the leading
 jet for the full
 (left) and high (right) $y$ range. Both jets lie in a central rapidity
 range $0 < \eta_{1,2} < 1$. The cross sections fall steeply with $E_T$
 which leads to increasing statistical errors. All three calculations agree
 within the statistical accuracy which is about $\pm 1$\% at low $E_T$ for the
 full $y$ range and $\pm 2$\% for the high $y$ range.}
 \label{fig:et}
\end{figure}

In Fig.\ \ref{fig:et} we compare the theoretical predictions for the NLO dijet
cross section as a function of the transverse energy $E_T$ of leading jet
with both jets at central 
pseudorapidities $0 < \eta_{1,2} < 1$. The three calculations agree within
the statistical accuracy. The errors are comparable for all three calculations.
They are about $\pm 1$\% ($\pm 2$\%) at low $E_T$ in the full (high) $y$
regime and larger at high $E_T$ due to the steeply falling cross section.

\begin{figure}[ttt]
 \begin{center}
  \vspace*{-1cm}
  \epsfig{file=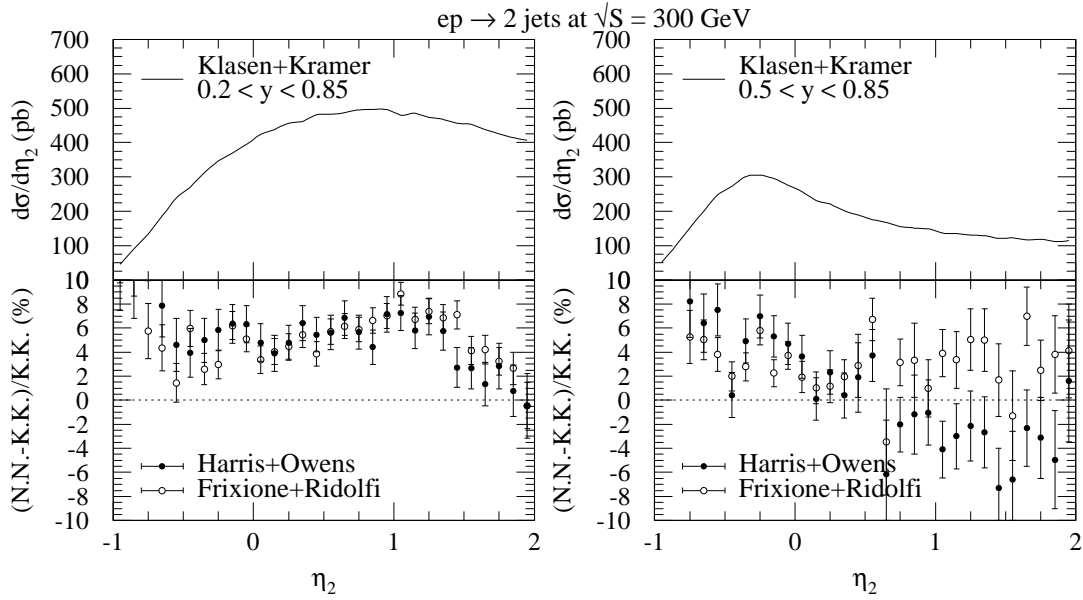,width=16cm}
  \vspace*{-1cm}
 \end{center}
 \caption{Comparison of three theoretical predictions for the NLO dijet
 cross section as a function of rapidity $\eta_2$ for the full
 (left) and high (right) $y$ range. The other jet lies in a central rapidity
 range of $0 < \eta_1 < 1$.}
 \label{fig:eta}
\end{figure}

In Fig.\ \ref{fig:eta} we present rapidity distributions for the NLO dijet
cross section with a central first jet $\eta_1 \in [0;1]$. HO
agree with FR for the full $y$ range and are about 5\% higher
than KK. In the high $y$ range, FR are about
4\% higher than KK, whereas HO have a slope
from +4\% in the backward direction to -4\% in the forward direction.
Studies have shown that HO agree with KK
very well for the resolved processes and for the Born and virtual
direct processes. This indicates that the difference, which is still under
study, may come from the real direct processes.
Within the statistical accuracy of about $\pm 2$\% the overall
agreement is, however, still acceptable.



\section{Remaining Uncertainties}
\label{sec:5}

The main remaining theoretical uncertainties arise from the dependence
of the hadronic cross section on the renormalization and factorization
scale $\mu$. This scale dependence is an artifact of the truncation of
the perturbative series at next-to-leading order. The scale $\mu$ has
to be larger than ${\cal O} (\Lambda_{\rm QCD})$ to ensure the applicability
of perturbation theory. Although the scale $\mu$ is in principle arbitrary
and the renormalization and factorization scales need not be equal,
the logarithmic NLO corrections can be made small by choosing a common scale
$\mu$ of the order of the hard scattering parameter. In jet photoproduction,
the relevant large scales are the transverse energies of the jets $E_{T_i}$
which need not be equal in NLO QCD. This justifies the choice of
$\mu = \max (E_{T_i})$ which we have used consistently throughout
this paper.

\begin{figure}[ttt]
 \begin{center}
  \vspace*{-1cm}
  \epsfig{file=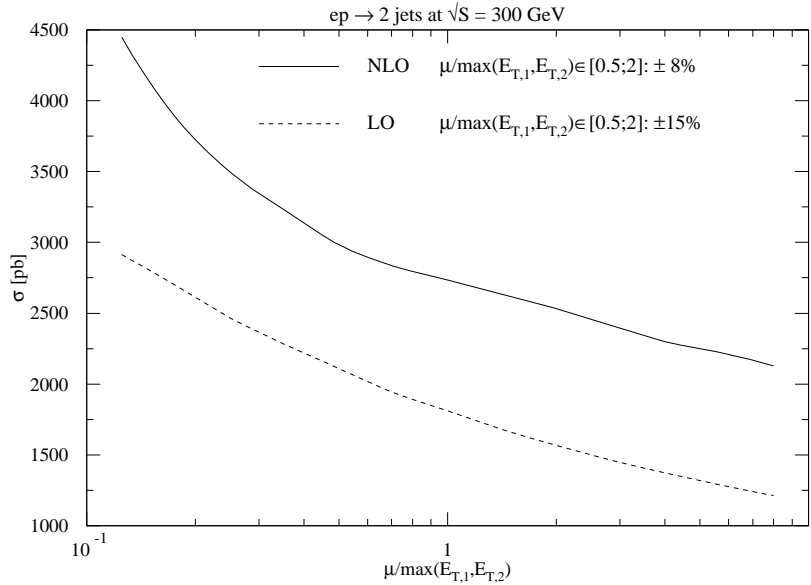,width=12cm}
  \vspace*{-1cm}
 \end{center}
 \caption{Dependence of the total dijet cross section on the common
 renormalization and factorization scale $\mu$. The NLO dependence is reduced
 with respect to the LO dependence, but is still not negligible.}
 \label{fig:mu2jet}
\end{figure}

It is customary to estimate the theoretical uncertainty of perturbative
calculations by varying $\mu$ around the central scale. The dependence of
the total dijet cross section with the same kinematic cuts as before on the
scale $\mu$ is plotted in Fig.\ \ref{fig:mu2jet}. 
We have checked that the calculations by HO and KK agree very well.
The LO cross section depends strongly and logarithmically on $\mu$ through the
strong coupling constant $\alpha_s(\mu)$ and the parton densities in the
photon $f_{q,g}^{\gamma}(x_{\gamma},\mu^2)$ and proton $f_{q,g}^{p}
(x_p,\mu^2)$. The dependence is reduced in NLO due to explicit logarithms in
the virtual and real corrections. However, it still amounts to a considerable
uncertainty of about $\pm 8$\%
which can be traced back to the photon factorization
scale dependence of the NLO resolved contribution. Whereas the LO photon
factorization scale dependence is almost completely cancelled by the NLO direct
contribution, the same cancellation for the NLO resolved contribution would
require the next-to-next-to-leading order (NNLO) direct contribution which is
unknown. The three-jet cross section is only accurate to LO QCD and suffers
from even larger scale uncertainties. They have been estimated to be about a
factor of two \cite{Klasen:1998cw}.

Further uncertainties arise from the power corrections in the
Weizs\"acker-Williams approximation \cite{Frixione:1993yw}. The non-logarithmic
terms have been included in all of our numerical results. Although power
corrections of ${\cal O}(m_e^2/Q^2)$ could be
expected to be negligible, an omission
of these terms results in an increase in the dijet and three-jet cross
section of about 5\%. The remaining uncertainty beyond this
${\cal O}(m_e^2/Q^2)$
correction is of ${\cal O}(\theta_e^2,m_e^2/E_e^2)$ and thus small.

While
theoretical calculations are on the parton level, experiments measure hadronic
jets. For LO three jet cross sections, every parton corresponds to a jet,
making
it impossible to implement an experimental jet definition in the theoretical
calculation. For NLO dijet cross sections, every jet consists of one or two
partons, and a jet definition can be implemented. The cone algorithm suffers
from uncertainties with $R_{\rm sep}$, which are absent in the $k_T$
algorithm used here \cite{Butterworth:1996ey}.

\begin{figure}[ttt]
 \begin{center}
  \vspace*{-1cm}
  \epsfig{file=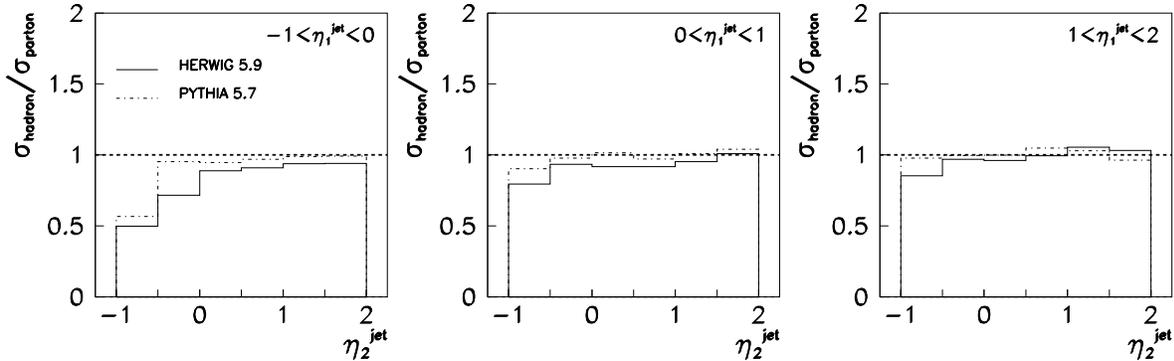,width=16cm}
  \vspace*{-1cm}
 \end{center}
 \caption{The ratio between the dijet cross section based on
 hadrons and that based on partons as predicted by HERWIG 5.9 and PYTHIA 5.7. 
 The ratio is given for the cross section as a function of the
 pseudorapidity of one of the jets while the other jet is restricted to 
 the pseudorapidity range, indicated in the figure.}
 \label{fig:hadronisation}
\end{figure}

Although jet cross sections are mainly sensitive to the dynamics of the
hard subprocess, the measured cross sections will at some level be 
effected by hadronization. These effects are expected to become 
smaller when the cross section refers to higher transverse energy jets.
We have estimated hadronization effects based on the leading order
Monte Carlo models HERWIG 5.9 \cite{Marchesini:1991ch} and PYTHIA
5.7 \cite{Bengtsson:1987kr}.
The jet cross section for hadrons in the final state was 
compared to the cross section of the partons produced from leading 
order matrix elements and parton showers (see Fig.\ \ref{fig:hadronisation}). 
In HERWIG the change in the cross section due to fragmentation 
was found to be less than 10\% in most of the kinematic regime. Only for 
events with one or more very backward jets ($\eta^{jet}<-0.5$) was a more 
sizeable change observed. For these events the cross section is reduced 
by up to 40\% due to fragmentation. In PYTHIA the reduction of 
the cross section is much smaller, but shows the same trend.
In a related study, presented in \cite{Harris:1997hz},
HERWIG 5.9 was used to compare the cross 
section for the final state hadrons to that for the partons of 
the leading order matrix elements. The relative difference between these 
cross sections was found to be less than 20\%, except again for events 
with very backward jets ($\eta^{jet}<-0.5$), where the change in the 
cross section can be as large as 50\%.
For the three jet measurement, a study of fragmentation effects using
PYTHIA 5.7 was presented in \cite{Strickland:1998}.
The three jet cross section based on the hadrons in the final state was 
compared to that of the partons produced in the hard subprocess and the 
parton showers. The cross section for hadrons was found to be 
approximately 5\% lower than that for partons.

The experimental uncertainty on the dijet cross section is dominated by 
systematic uncertainties up to transverse jet energies of approximately 
25 GeV, depending on the angles of the jets.  
At higher transverse energies statistical uncertainties dominate. The 
systematic uncertainties are roughly between 10 and 20\% \cite{dijet}.
The experimental uncertainty in the three jet measurement is dominated by 
systematics up to a three jet mass of approximately $100$~GeV and 
statistics dominated at higher masses. Here, the systematic uncertainties 
are of the order of 20\% \cite{Breitweg:1998uv,Strickland:1998}.

These measurements correspond to luminosities of $6.3$ and 
$16$~pb$^{-1}$, respectively. Up to the beginning of 1999 the HERA 
experiments have each collected around $50$~pb$^{-1}$. 
When these data are used to repeat the discussed measurements,
it will be possible to reduce the statistical uncertainties significantly
and to extend the measurement to higher transverse energies and masses. 
Moreover, it is likely that the increase in statistics can be exploited  
to reduce the systematic uncertainties as well.
For the dijet analysis, it was estimated that, when using all available 
data, statistical uncertainties should dominate the
measurement only above transverse energies of approximately 50 GeV, 
again depending on the angles of the jets. 
In the long term, after the luminosity upgrade planned in the year 2000, 
HERA is aiming to deliver about $250$~pb$^{-1}$ of luminosity each year. 
This will allow for the measurement of jet photoproduction cross sections 
up to still higher transverse energies and masses. 



\section{Conclusions}
\label{sec:6}

We have presented a detailed comparison of three theoretical predictions
for LO three-jet and NLO dijet cross sections as measured recently by
ZEUS. We found that in general all three calculations agree within the
statistical accuracy of the Monte Carlo integration. In certain restricted
regions of phase space, the calculations differ by up to 5\%. We briefly
discussed remaining theoretical and experimental uncertainties and future
developments.


\section*{Acknowledgments}

Work in the High Energy Physics Division at Argonne National Laboratory is
supported by the U.S. Department of Energy, Division of High Energy Physics,
under Contract W-31-109-ENG-38.  Work at NIKHEF
is supported by the Dutch Foundation for Research on Matter (FOM).
The FR results were produced by us, and we thank Stefano Frixione for
assistance with his computer code.




\end{document}